\documentclass[aps,pre,onecolumn,groupedaddress,showpacs]{revtex4}

\usepackage{amssymb,amsmath,amsfonts,graphicx}
\usepackage{bm}

\begin{document}

\title{Softening of edges of solids by surface tension}
\author{Serge Mora}
\email[]{smora@univ-montp2.fr}
\affiliation{Laboratoire de M\'ecanique et de G\'enie Civil de Montpellier. UMR 5508, Universit\'e Montpellier 2 and CNRS. Place Eug\`ene Bataillon. F-34095 Montpellier Cedex, France.}
\author{Yves Pomeau}
\affiliation{University of Arizona, Department of Mathematics, Tucson, USA.}

\date{\today }

\begin{abstract}
\textbf{Abstract}
Surface tension tends to minimize the area of interfaces between pieces of matter in different thermodynamic phases, be they in the solid or the liquid state. This can be relevant for the macroscopic shape of very soft solids, and lead to a roughening of initially sharp edges. We calculate this effect for a neo-Hookean elastic solid, with assumptions corresponding to actual experiments, namely the case where an initially sharp edge is rounded by the effect of surface tension felt when the fluid surrounding the soft solid (and so surface tension) is changed at the solid/liquid boundary. We consider two opposite limits where the analysis can be carried to the end, the one of a shallow angle and the one of a very sharp angle. Both cases yield a discontinuity of curvature in the state with surface tension although the initial state had a discontinuous slope.  
\end{abstract}
\maketitle
Because of surface tension, the area of a droplet of liquid tends to a minimum at a given volume. This yields a sphere if other effects like gravity or interaction with a solid surface are negligible or absent. On the contrary, in solids, usually, one does not consider surface tension  because its relative strength compared to elastic forces is ruled by a very small length, called sometime the elasto-capillary length equal to the ratio $\frac{\gamma}{\mu}$, $\gamma$ capillary constant (or the surface energy) and $\mu$ elastic shear modulus. For ordinary solids, this length is comparable to atomic scales or even smaller and so cannot be seen as pertinent at macroscales. However, in some soft materials, this elasto-capillary length may be much bigger than molecular sizes and may even be within the macroscopic domain, in the millimetric range. These large values come from the decoupling between the origin of the shear modulus and the origin of surface energy. Note that the surface energy is not always equivalent to the surface tension for solids. The surface energy gives the work needed to create additional surface area by cleaving, while the surface tension gives the work needed to create additional surface area by stretching \cite{Shuttleworth1950,Cammarata1994,Muller2004}. For liquids they are equal but, for solids they may even have opposite signs. However, for soft solids like hydrogels having a macroscopic elasto-capillary lengths, the surface tension and the surface energy are equivalent. Hereafter, we will consider only this kind of materials and then it is not necessary to distinguish surface tension from surface energy.

\begin{figure}[!h]
\includegraphics[width=0.6\textwidth]{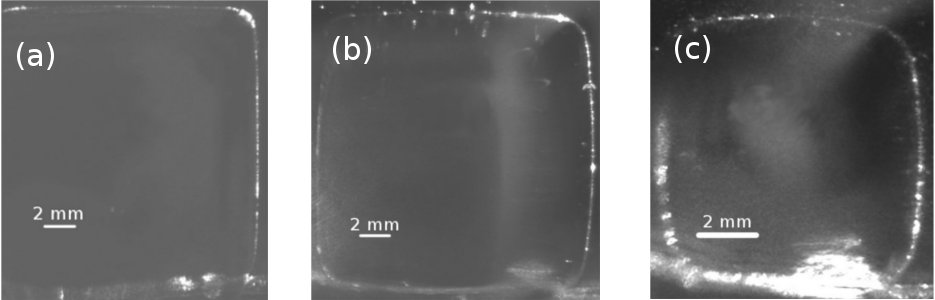}
\caption{Capillary deformation of elastic rods (hydrogel) after being released from molds having initially square cross-sections, and immersed into silicon oil.  The surface tension of the solid-liquid interface is 36.5 mN/m. The shear moduli of the gels are (a) 125 Pa, (b) 60 Pa, (c) 60 Pa. The initial size of the cross section is (a) 12 mm, (b) 12 mm, (c) 6 mm. The cross-section is illuminated by a laser sheet and it is observed with a slight angle to avoid optical distortions. The top right boundary of the cross-section appears here in white, without distortion. The initially square cross-sections are rounded. This rounding effect is more pronounced as the ratio of the elasto-capillary length to the initial size of the cross-section is larger. Transferring the rods from oil into water, thereby suppressing the surface energy, the original square cross-sections is recovered (not shown here) \cite{Mora2013}.}
\label{fig : manips}
\end{figure}

A rather spectacular occurrence of capillarity in elastic (soft) solid is the spontaneous growth of undulations on cylinders, by the same mechanism as the Rayleigh-Plateau instability of columns of liquids \cite{mora_prl2010,Henann2014}. Contrary to the case of liquids, this instability has been found to occur beyond a threshold. Surface tension driven deformations of slender elastic solids also occur below this threshold, depending on the shape of its cross section: a circular elastic cylinder shortens in the longitudinal direction and stretches transversely and the sharp edges of a square based prism get rounded off as its cross sections tend to become circular (Fig. \ref{fig : manips}) \cite{Mora2013}. This rounding effect is related to the shape change of periodic ridge surface profiles in hydrogel resulting from deformation driven by their surface tension \cite{Jagota2012,Jagota2014}, and it has been investigated through numerical simulations \cite{Jagota2002,Jagota2013}. Surface tension of solids has also been found to be a barrier for some instabilities like creasing, wrinkling, or cavitation in soft elastic materials \cite{Mora_biot2011,Chen2012,Crosby2011,Benamar2010,Benamar2011}. Another recently observed effect of surface tension on the equilibrium shape of soft elastic solids is the deformation close to a contact line. It has been found to be determined by a balance of the three interfacial tensions at the contact line \cite{Dufresne2012b,Dufresne2013,Style2013,Jagota2013b}. In that case, the driving force is the surface tension of the third phase: without it, no deformation would occur. 

Here, we focus on deformations that are induced only by the surface tension of the deformed solid, without any other forces (except the elastic restoring forces). This is in contrast with the contact line problem, or the stabilization of instabilities. Because of the existence of an intrinsic length scales, capillary phenomena in soft solids are more complex to analyse than in liquids. Like a liquid a solid tends to deform to minimize its free surface under the effect of surface tension, but this is opposed by the ensuing deformation of the solid. Therefore the final equilibrium state of deformation and the shape of the free surface result from a balance between elastic and capillary forces, usually quite complex. The present work is devoted to a study of the roughening of soft solids by capillary forces. We consider a solid filling initially ({\it{ie}} without surface tension) a wedge of arbitrary angle, and later left to the opposing effects of surface tension and of elastic forces. Because the initial geometry is free of any length scale, the final shape should be unique, up to a dilation factor depending on the elasto-capillary length and on the given initial angle of the wedge. However, things are slightly more complex because the deformation induced by the capillary forces diverges logarithmically for an unbounded wedge. 

Because there is no small parameter in the problem, one expects the deformation induced by surface tension to be of finite amplitude, at least in some places. This requires to use elasticity for finite strains, outside of the range of validity of the usual Hookean approximation of a linear relationship between strain and stress. In the case we consider with finite strain, we use the neo-Hookean elasticity, known to describe fairly well the soft solids, or hydrogels, used for experiments. 

One can think to two versions of the wedge problem: it can be a 2D wedge bounded by two half planes making a prescribed angle at infinity, or it can be  a 3D wedge such that the elastic material tends at infinity to a cone of given aperture. We consider below the 2D case only, leaving the elastic cone  to a future publication. In the formulation of the problem, one parameter only is left, namely the magnitude of the angle at infinity, be it the angle between the two planes in the 2D wedge or, in 3D, the aperture of the cone, if this one has a circular basis. Because of the geometry one can think to two limits for an analytical  approach: an almost flat wedge, where the angular deviation from a flat surface  is small, and the opposite limit of a sharp wedge where the angle at infinity is very small. Both cases can be treated analytically, with different methods of course. We expose below the corresponding calculations. In the case of a very sharp wedge, not surprisingly we have to use a form of lubrication approximation, that is to assume a much faster dependence on the coordinate perpendicular to the axis of symmetry than parallel to it. In the case of an almost flat wedge instead the capillary force, which is proportional to the small angle at the tip is small so that a limit of small strain can be used, and the problem can be dealt with by using linear perturbations. This makes the problem solvable in principle by using Fourier transform, but with some care because of divergences due to the large distance behavior of the perturbation induced by surface tension. 

We outline first the equations of neo-Hookean elasticity with surface tension included (section \ref{sec:deriv+ST}). We use later in section 
\ref{sec:wedge}  the linearized Hookean form of those equations to derive the deformation induced by surface tension on a shallow wedge. Then, in section \ref{sec:sharp} we solve the same problem with a very sharp edge in the rest state. It ends with a concluding section. 
  
\section{Derivation of the equations and boundary condition with surface tension}
\label{sec:deriv+ST}
\subsection{Cauchy-Poisson equations}
We consider the smoothening of a wedge under the effect of surface tension of elastic materials which 

{\it{i)}} can stand large amplitude deformations while remaining in a reversible state, 

{\it{ii)}} are incompressible, a condition accounted for via a Lagrange multiplier in the equations,  

{\it{iii)}} have a macroscopic "elasto-capillary length", making surface tension as well as Hookean or neo-Hookean elasticity relevant.

\begin{figure}[!h]
\includegraphics[width=0.35\textwidth]{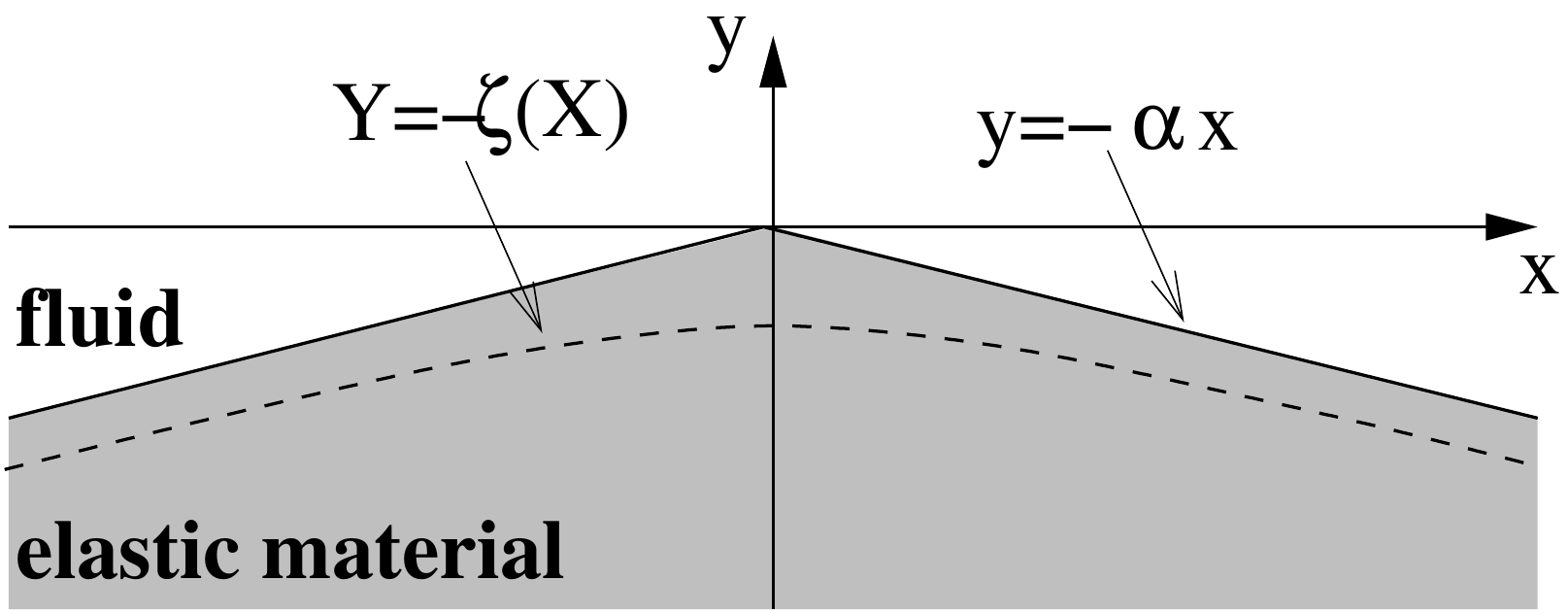}
\caption{Schematic view of a bi-dimensional wedge initially bounded by two half planes.   The solid line is the boundary of the elastic medium without surface tension (unperturbed state). It gets deformed upon an increase of the surface tension (dashed line).}
\label{fig : angle}
\end{figure}

The unperturbed state is the state without surface tension, namely the elastic solid fills the wedge $ y < - \alpha |x|$ where $\alpha$ is a given number and $(x,y)$ the regular Cartesian coordinates in the unperturbed state (Fig. \ref{fig : angle}). We expect that, because of surface tension, the wedge becomes rounded so that the surface of the solid becomes a smooth curve of Cartesian equation $ Y = \zeta(X) $ with $\zeta(X) \rightarrow - \alpha |X|$ as $|X| \rightarrow  \infty$. This last condition expresses that, far from the wedge, surface tension does not generate any Laplace's pressure on a flat surface, $(X,Y)$ being the positions in the strained state.  

Because we have to deal with finite strain (instead of the usual small strain limit of elastic theory, or Hookean limit), we need to formalize more than done usually the derivation of the equations of elasticity. We limit our exposition to the 2D case. The deformation is characterized by a map from the undisturbed state with coordinates $(x,y)$ to a disturbed state ${\bf{R}}(x,y)= (X(x, y), Y(x, y))$ (boldface being here and thereafter for vectors). In this Lagrangian framework, the actual location in space is parametrized by the coordinates of the preimage in the undisturbed (rest) state. The boundary condition (b.c.) on the surface is the condition that the pair $(X_{\partial} (x) = X(x, -  \alpha |x|), (Y_{\partial} (x) = Y(x,  - \alpha |x|))$ is a parametric representation of the curve of equation $ Y_{\partial} = \zeta(X_{\partial})$, the subscript $\partial$ being to indicate that a quantity is evaluated on the surface of the wedge (here a line). 

Incompressibility is imposed by writing that the determinant of the first derivatives of ${\bf{R}}(x,y)$ is equal to one, namely that $X_{,x} Y_{,y} - X_{,y} Y_{,x} = 1$,
where $X_{,x}$ is for $\frac{\partial X}{\partial x}$, etc. The elastic energy is a function of the strain. The strain tensor $\omega_{ij}$ is defined by writing the change of square length of an infinitesimal displacement as $$ ds'^2 - ds^2 = \sum_{ij} \omega _{ij} dx_i dx_j \mathrm{.}$$ The convention of summation (without symbol $\sum$) on like indices (here indices $i$ and $j$) shall be assumed in the rest of this paper. As well-known $
\omega _{ij} = R_{k,j} R_{k,j} - \delta_{ij}$
where $R_{k,i}$ is for $\frac{\partial R_k}{\partial x_i}$, $R_k$ being the $k$-component of the displacement (i.e the vector of components $(X,Y)$) and $x_i$ is the Cartesian coordinate of position in the rest state. Lastly $\delta_{ij}$ is the discrete Kronecker symbol.  In 2D the components of the strain are $$\omega_{xx} = (X_{,x})^2 + (Y_{,x})^2 - 1 \mathrm{,}$$  $$\omega _{yy} = (X_{,y})^2 + (Y_{,y})^2- 1 \mathrm{,}$$ and $$\omega_{xy} = \omega _{yx} = X_{,y} X_{,x} +  Y_{,y} Y_{,x} \mathrm{.}$$
 The relation between displacement and strain is nonlinear to insure that a solid body rotation does not induce any strain, whatever is its amplitude. Otherwise a rotation of angle $\theta$ leaves a non zero contribution to the part of $\omega _{xx}$ linear with respect to the displacements, proportional to $(1 - \cos(\theta))$. Notice too the underlying assumption that, even for the finite deformations we consider, the solid in the deformed state does not cross itself. This is in principle enforced by imposing that the mapping from $(X(x,y), Y(x,y))$ to $(x,y)$ and its inverse are one to one. 

The elastic energy is a function of the strain tensor just defined. In the absence of any preferred direction it must depend on one of the two scalars (i.e. invariants under global rotation) that can be made out of the strain tensor. These are the sum of the square of its components, namely $\omega _{ij}  \omega _{ij}$ and the trace, $\omega_{ll}$. There is a wide range of possibilities \cite{Ogden1984} for the explicit dependence of the energy with respect to the two invariants (the third invariant, the determinant, cannot enter into the energy for an incompressible material because it is constant). We assume that the elastic energy is proportional to the trace itself, which defines a neo-Hookean solid. With this choice, the quantity to be minimized (called later the energy, although it includes both an elastic energy {\it{stricto sensu}} and a Lagrange term to ensure incompressibility)  with respect to small variations of the displacement  reads (in 2D): 
\begin{equation}
\mathcal{E} = \int \mathrm{d}x \int \mathrm{d}y \mathcal{W}
\textrm{,}
\label{eq:functionalI}
\end{equation}
where 
\begin{equation}
\mathcal{W} =  \frac{\mu}{2} \left( X_{,x}^2 + Y_{,x}^2 + X_{,y}^2 + Y_{,y}^2- 2 \right) - q(x,y) (X_{,x} Y_{,y} - X_{,y} Y_{,x} )
\textrm{.}
\label{eq:int}
\end{equation}
To make the material stable 
the shear modulus $\mu$ must be positive. The Lagrange multiplier $q(x,y)$ allows to impose the condition $X_{,x} Y_{,y} - X_{,y} Y_{,x} = 1$ everywhere. The variation of $\mathcal{E}$ with respect to $X(.)$ and $Y(.)$ yields the equilibrium equations.  A small change $(\delta X(x,y), \delta Y(x,y))$ ( $\delta R_i (x_j)$ in intrinsic notations) of the displacement yields a change of $\mathcal{W} $ that can be written as:
\begin{equation}
\delta {\mathcal{W}} =  \frac{\partial \delta R_i (x,y)}{\partial x_j} \frac{\partial \mathcal{W}}{\partial R_{i,j}}  
\textrm{.}
\label{eq:intvar}
\end{equation}
Let us define the stress tensor as $\sigma_{ij} = \frac{\partial {\mathcal{W}}}{\partial R_{i,j}} $ (in our notations, $R_{i,j} = \frac{\partial R_i}{\partial x_j}$). The Euler-Lagrange equations of equilibrium read:
\begin{equation}
\frac{\partial \sigma_{xx} }{\partial x} + \frac{\partial \sigma _{xy}}{\partial y} = 0
\textrm{,}
\label{eq:Eulintx}
\end{equation}
  and 
\begin{equation}
\frac{\partial \sigma_{yy}}{\partial y} + \frac{\partial \sigma_{xy}}{\partial x} = 0
\textrm{.}
\label{eq:Eulinty}
\end{equation}
We shall refer below to those equations as the Cauchy-Poisson equations (they are general, independent on the particular form of ${\mathcal{W}}$ as a function of the strain).  In the neo-Hookean model, the components of the stress are
$ \sigma_{xx} =  \mu X_{,x} - q Y_{,y} 
\mathrm{;}$ 
$ \sigma_{yy} =   \mu Y_{,y} - q X_{,x}
 \mathrm{;}$ 
$ \sigma_{xy} =   \mu X_{,y}  + q Y_{,x} 
\mathrm{;}$ 
and 
$ \sigma_{yx} =   \mu Y_{,x} + q X_{,y} 
\mathrm{.}$ 
Note that, for finite perturbations the stress tensor so defined is not symmetric.  The Cauchy-Poisson equations read explicitly:
\begin{equation}
\mu \nabla^2 X + q_{,y} Y_{,x} -  q_{,x} Y_{,y} = 0
\textrm{,}
\label{eq:CPx}
\end{equation} 
and 
\begin{equation}
\mu \nabla^2 Y + q_{,x} X_{,y} -  q_{,y} X_{,x} = 0
\textrm{.}
\label{eq:CPy}
\end{equation} 

\subsection{Boundary conditions and the value of $q$}
\label{susec:Biotbc}
Thanks to our Lagrangian variables, the b.c. are imposed on the free surface in the {\it{undisturbed/rest}} state. 
They follow from the boundary terms leftover in the integration by part leading to equations (\ref{eq:Eulintx}-\ref{eq:Eulinty}) and similar ones for the other components of the displacement. The b.c.'s on a boundary of Cartesian equation $ y = y(x)$ at rest are derived as follows. One writes first the energy as 
$$ \mathcal{E} = \int_{-\infty}^{\infty}\mathrm{d}x \int_{-\infty}^{y(x)} \mathrm{d}y   \left(\frac{\mu}{2} \left( X_{,x}^2 + Y_{,x}^2 + X_{,y}^2 + Y_{,y}^2- 2 \right) - q(x,y) (X_{,x} Y_{,y} - X_{,y} Y_{,x} ) \right) \mathrm{.}$$ Some contributions to the variation of this energy due to a variation $\delta X(x,y)$ of $X(x,y)$ read:

$$ \delta \mathcal{E} =\mu  \int_{-\infty}^{\infty}\mathrm{d}x \int_{-\infty}^{y(x)} \mathrm{d}y \left( \delta X_{,x} X_{,x} + \delta X_{,y} X_{,y} \right)  \mathrm{.}$$ 
From the identity $$\int_{-\infty}^{\infty}\mathrm{d}x \int_{-\infty}^{y(x)}  \mathrm{d}y F_{,x}(x,y) = \left[  \int_{-\infty}^{y(x)}  \mathrm{d}y F(x,y) \right]_{x=-\infty}^{x = +\infty} + \int_{-\infty}^{\infty}\mathrm{d}x (-y_{,x}) F(x,y(x))  \mathrm{,}$$ one can derive the boundary terms leftover in the integrations by part leading to the Cauchy-Poisson equations (they arise  from the last term in the identity above). Those leftover terms are proportional to the value of $\delta X$ on the line $ y = y(x)$ and must vanish to make stationary  the elastic energy. The identity above is used by writing $\delta X_{,x} X_{,x} = (\delta X X_{,x})_{,x} -  X_{,xx} \delta X$. That the contribution of the last term vanishes is ensured by the Cauchy-Poisson equation, although the first term yields a boundary term on the line $ y = y(x)$ that must be cancelled independently of the bulk equations. 

The final result are two b.c., one coming from the term proportional to $\delta X$, the other from terms proportional to $\delta Y$ on the boundary. In the variation $\delta  \mathcal{E}$ those b.c. express the cancellation of integrals over $x$ only. They read explicitly:
\begin{equation}
\mu( X_{,y} - y_{,x} X_{.x} ) + q Y_{,x} +  y_{,x} q Y_{,y}  = 0
\textrm{,}
\label{eq:Eulbcexpl1}
\end{equation}
and 
\begin{equation}
\mu( Y_{,y} - y_{,x} Y_{,x} ) - q X_{,x} -  y_{,x}  q X_{,y}  = 0
\textrm{.}
\label{eq:Eulbcexpl2}
\end{equation}
All conditions are imposed at $ y = y(x)$ and for any value of $x$. Those b.c. can also be written in terms of the stress tensor. The first one reads $\sigma_{xy} - y_{,x} \sigma_{xx} = 0$ and the other $\sigma_{yy} - y_{,x} \sigma_{yx} = 0 \mathrm{.}$ 
\subsection{Surface tension}
\label{susec:surften}

Consider an arbitrary free surface at rest. As we plan to introduce the effect of surface tension, we have to define more precisely what is meant by "at rest". We assume (something corresponding to the way the experiments are done) that the rest state is without surface tension. Therefore the surface tension changes already the strain in the solid, unless the free surface is flat and does not induce any Laplace stress \cite{Gurtin1975,Gurtin1978}. 

Laplace's pressure comes from the variation of the capillary energy. For the materials we consider, the capillary energy is proportional to the area times the capillary constant $\gamma$.  With our choice of Lagrange parametrization of the surface, this capillary energy reads:
$$ {\mathcal{E}}_{cap} = \gamma \int {\mathrm{d}}x \sqrt{X_{\partial, x}^2 + Y_{\partial, x}^2} \mathrm{,}$$ where $ X_{\partial, x} =\frac{{\mathrm{d}}X_{\partial}}{{\mathrm{d}}x} = X_{,x} + y_{,x} X_{,y}$, $ y = y(x)$ being the Cartesian equation of the free surface in the rest state. 

The capillary energy adds a contribution to the value of the stress on the surface. With this contribution of the surface energy, the b.c. on the (arbitrary) free surface read now:
\begin{equation}
\sigma_{xy} - y_{,x} \sigma_{xx} = \gamma \left(\frac{X_{\partial, x}}{\sqrt{X_{\partial, x}^2 + Y_{\partial, x}^2}} \right)_{\partial, x} 
\mathrm{,}
\label{eq:bcLap1}
\end{equation}
and 
\begin{equation}
\sigma_{yy} - y_{,x} \sigma_{yx} = \gamma \left(\frac{Y_{\partial, x}}{\sqrt{X_{\partial, x}^2 + Y_{\partial, x}^2}} \right)_{\partial, x} 
\mathrm{.}
\label{eq:bcLap2}
\end{equation}
We consider in the next section an application of this set of equations, namely the change of shape induced by surface tension on a wedge with a given angle at rest. 

 \section{The wedge problem}
 \label{sec:wedge}

We consider the smoothening of a wedge of elastic material under the effect of surface tension.  In 2 dimensions, the unperturbed state is the state without surface tension, namely the elastic solid fills the wedge $ y < - \alpha |x|$ where $\alpha$ is a given number. Below we look at the two limits, one after the other:  the limit of a very shallow angle, i.e, at the limiting case $\alpha$ close to zero and to the limit of a very sharp angle, i.e. $\alpha$ very large. The general case, namely the arbitrary angle for the undeformed tip, will be dealt with (mostly numerically) in a companion paper. 

\subsection{$\alpha$ small}
\label{subsec:alphasmall}
In this limit,  we shall assume (and this has to be checked at the end) that the strain is small, so that one may restrict oneself to the equations of linear Hookean elasticity. Let $ X = x + u(x,y)$ and $Y = y + v(x,y)$ be the two Cartesian components of the displacement, with the assumption that the derivatives of $u(.)$ and $v(.)$ are small (i.e. that the strain is small). This leads to neglect everywhere nonlinear terms in the equations of elasticity. From incompressibility $u_{,x} + v_{,y} = 0 \mathrm{.}$  The linear part of the Cauchy-Poisson equations read:
\begin{equation}
 \mu \nabla^2 u -  q_{,x}  = 0 
 \mathrm{,}
 \label{eq:Cauchy12DX}
 \end{equation}
  and 
\begin{equation}
 \mu \nabla^2 v -  q_{,y}  = 0 
 \mathrm{.}
 \label{eq:Cauchy12DY}
 \end{equation}
To satisfy the incompressibility condition we introduce the function $\Psi(x,y)$ such that $ v = \frac{\partial \Psi}{\partial x}$ and $ u  = - \frac{\partial \Psi}{\partial y}$. The Cauchy-Poisson condition becomes $ \Delta^2 \Psi = 0 \mathrm{.}$  
In the b.c.  one neglects the nonlinear terms and takes $y_{,x}$ as negligible, but not $y_{,xx}$ which is a Dirac delta function. 
To get the explicit form of the b.c. one uses the leading order approximation of $X_{{\partial},x}$ and of $Y_{\partial, x}$: $$X_{{\partial},x} = X_{,x} + y_{,x} X_{,y} \approx  X_{,x}  \approx 1 + u_{,x} \mathrm{,}$$ and $$Y_{\partial, x} = Y_{,x} +  y_{,x} Y_{,y} \approx v_{,x} + y_{,x} \mathrm{.}$$ Therefore $$  \left(\frac{X_{\partial, x}}{\sqrt{X_{\partial, x}^2 + Y_{\partial, x}^2}} \right)_{\partial, x} \approx u_{,xx} \mathrm{,}$$ and:  
$$\left(\frac{Y_{\partial, x}}{\sqrt{X_{\partial, x}^2 + Y_{\partial, x}^2}} \right)_{\partial, x}  \approx v_{,xx} + y_{,xx}\mathrm{.}$$ The second derivative $y_{,xx}$ is a Dirac delta function, equal to $ y_{,xx} = - 2 \alpha \delta(x) \mathrm{.}$ The writing of the b.c. is done by considering $y_{,x}$ as small and also by taking into account that $q = \mu$ in the rest state. Therefore the b.c. are:
\begin{equation}
 \mu (u_{,y} + v_{,x})  -\gamma u_{,xx} = 0 
 \mathrm{,}
 \label{eq:Cauchybcca1}
 \end{equation}
 and
 \begin{equation}
 \mu (v_{,y} - u_{,x}) -\gamma (v_{,xx} - 2 \alpha \delta(x)) - (q - \mu)= 0 
 \mathrm{.}
 \label{eq:Cauchybcca2}
 \end{equation}
 In the limit $\alpha$ small one neglects at leading order the slope of the surface in the definition of the manifold where the b.c.'s are imposed. The b.c.'s are on the line $y=0$ and read as follows in terms of the function $\Psi$:
\begin{equation}
\mu (\Psi_{,yy} - \Psi_{,xx}) -\gamma \Psi_{,xxy}= 0 
 \mathrm{,}
 \label{eq:Cauchybcca1Psi}
 \end{equation}
\begin{equation}
2 \mu \Psi_{,xy} -\gamma( \Psi_{,xxx} - 2 \alpha \delta(x)) - (q - \mu)= 0 
 \mathrm{,}
 \label{eq:Cauchybcca2Psi}
 \end{equation}
 where $ (q - \mu) $, the contribution to $q$ proportional to $\alpha$, is derived as a function of $\Psi$ from $(q - \mu)_{,x} = - \mu \Delta \Psi_{,y}$, a consequence of equations (\ref{eq:Cauchy12DX}) and   (\ref{eq:Cauchy12DY}).  
 For $y$ negative the solution of the bi-harmonic equation $\Delta^2 \Psi = 0$ relevant for the present problem read in Fourier transform:
 $$\Psi(x,y) = \frac{1}{2\pi} \int_{-\infty}^{+\infty} {\mathrm{d}}k \ e^{ikx} e^{|k|y} [a(k) + b(k) y]\mathrm{.}$$ It decays to zero for $y$ large negative (or at least it does not grow faster than algebraically) and depends on two functions of $k$, $a(.)$ and $b(.)$ to be derived from the b.c.'s  written in Fourier transform. The b.c.'s yield:
 $$ a(k) = \frac{i\gamma \alpha}{k( 2 \mu |k| + \gamma k^2)} \mathrm{,}$$ 
 $$ b(k) = - |k| \frac{i\gamma \alpha}{k( 2 \mu |k| + \gamma k^2)} \mathrm{.}$$ 
 Because $a(k)$ and $b(k)$ are purely imaginary odd functions of $k$, the integral defining $\Psi$ becomes:
 $$\Psi(x,y) = \frac{\gamma \alpha}{\pi} \int_{0}^{+\infty} {\mathrm{d}}k \frac{\sin(kx)}{k} e^ {k y} \frac{(yk - 1)}{( 2 \mu k + \gamma k^2)}\mathrm{.}$$
This has a logarithmic divergence at $k =0$. However the $x-$ component of the displacement  $u = - \Psi_{,y}$ is given by an integral converging everywhere:
\begin{equation}
 u = - \frac{\gamma \alpha}{\pi} \int_{0}^{+\infty} {\mathrm{d}}k \frac{\sin(kx)}{k} e^ {k y} \frac{k y}{( 2 \mu + \gamma k)}\mathrm{.}
 \label{eq:uinteg}
 \end{equation} 
The behavior of $u(x,y)$ near $x = y =0$ is found by taking  $x = \rho \cos(\theta)$ and $y = \rho \sin(\theta)$, with $\theta$ fixed angle between $\pi$ and $2 \pi$ and $\rho$ tending to zero. Taking $\kappa = k\rho$ as integration variable, one finds:
$$ u \approx  - \frac{ \alpha y}{\pi} G(\theta)$$ where $$ G(\theta) = \int_{0}^{+\infty} {\mathrm{d}}\kappa \frac{\sin(\kappa \cos(\theta))}{\kappa} e^ {\kappa\sin(\theta)} \mathrm{.}$$ 

The numerical function $G(\theta)$ is well defined for any value of $\theta$ in the interval $[\pi, 2\pi]$.  
The large distance behavior of $u(x,y)$ follows from the estimation of the integral in equation (\ref{eq:uinteg}) at large $x$ and $y$. Taking $\kappa = k \rho$ as integration variable one obtains:
$$ u \approx -\frac{\gamma \alpha}{\mu} F(\theta) \mathrm{,}$$ where 
$$ F(\theta) =  \sin(\theta) \int_{0}^{+\infty} {\mathrm{d}}\kappa \sin(\kappa \cos(\theta)) e^ {\kappa\sin(\theta)} \mathrm{.}$$ 
The other component of the displacement, $v(x,y)$ cannot be computed directly from the integral expression of $\Psi$ because it diverges. To obtain it, let us solve first the incompressibility condition 
$v_{,y} + u_{,x} = 0$ by putting there the value of $u(.)$ just found, and by writing $v(x,y) = -\int_0^{y} {\mathrm{d}}y  u_{,x} + v(x, y=0) \mathrm{.}$

This yields:
 $$ v(x,y) - v(x, 0) =  \frac{\gamma \alpha}{\pi} \int_{0}^{+\infty} {\mathrm{d}}k \frac{\cos(kx)}{( 2 \mu + \gamma k)} \left( y e^{ky} -  \frac{1}{k}(e^ {k y} - 1) \right)\mathrm{.}$$ 
The integral defining $v(x,y) - v(x, 0) $ converges for any value of $x$ and $y$.
When using the relationship between $v$ and $\Psi$, one finds again a diverging integral for $v(x,0)$. However the $x$-derivative of $v(x,0)$ is given by the following converging integral:
\begin{equation}
v_{,x}(x,0) = \frac{\alpha\gamma}{\pi} \int_{0}^{+\infty} {\mathrm{d}}k \frac{\sin(kx)}{( 2 \mu + \gamma k)}\mathrm{.}
\label{eqn : vx}
\end{equation}
This can be integrated from $x = 0$ to an arbitrary value of $x$, to give:
\begin{equation}
v(x,0) - v(0,0)  = \frac{\alpha \gamma}{\pi} \int_{0}^{+\infty} {\mathrm{d}}k \frac{1 - \cos(kx)}{( 2 \mu k+ \gamma k^2)}\mathrm{,}
\label{eqn : v}
\end{equation} 
a converging integral. The value of $v(0,0)$ is arbitrary because the equations for the displacement, namely the equations of elasticity and the b.c., are unchanged by the addition of an arbitrary constant to $v(x,y)$. However if the length scale $d$ defining a typical macroscopic length-scale (it can be the lateral dimension of the planes forming the wedge) is introduced ($d \gg \gamma/\mu$), it defines a cutoff in the integral arising from $v=\frac{\partial \Psi}{\partial x}$:
$$
v(0,0)=\frac{2\alpha \gamma}{\pi}\int_{1/d}^{\infty}\frac{-1}{2\mu k+\gamma k^2} \mbox{d}k =-\frac{\alpha}{\pi}\frac{\gamma}{\mu}\ln\left(1+\frac{2\mu}{\gamma}d\right).
$$
Therefore, the displacement of the tip of the wedge is then $\delta \simeq \frac{\alpha}{\pi}\frac{\gamma}{\mu}\ln\left(\frac{2\mu d}{\gamma} \right).$\\

The displacement along $y$ is given by the following converging integral:
\begin{equation} 
v(x,y)-v(0,0) = \frac{\alpha \gamma}{\pi} \int_{0}^{+\infty}  \frac{{\mathrm{d}}k}{( 2 \mu k+ \gamma k^2)} \left( 1 + \cos(kx) e^{ky} (ky -1)\right)\mathrm{.}
\label{eq:vdefinitif}
\end{equation}
This shows an interesting behavior of the deformation at large distance from the wedge: it is of order $\alpha \gamma/\mu$ times a function of the direction, namely a function of $y/x$. The same is true for the horizontal component of the displacement as well.  This is in agreement with the property that the elastic force across a large circle centered on the wedge should balance the capillary force coming from the neighbourhood of the wedge: at large distance, because the displacement is a function of the angle only, the strain decays like $1/r$, $r$ distance to the wedge. Therefore the stress decays also like $1/r$ and, once integrated over a large circle to give the total force, this yields a constant, independent on $r$. Moreover this force is proportional to $\gamma \alpha$, the order of magnitude of the total capillary force generated by a change of orientation of the surface by an angle of order $\alpha$. \\

The shape of the surface near $x=0$ is found by adding the base solution $ y = - \alpha|x|$ and the perturbation $v(x,y=0)$ just derived. 
$v_x(x,0)$ can be calculated explicitly from  Eq. \ref{eqn : vx}:
$$
v_{,x}(x,0)=\frac{2\alpha}{\pi}\left[\mbox{ci}\left(\frac{2\mu}{\gamma}x\right)\sin\left(\frac{2\mu}{\gamma}x\right)-\mbox{si}\left(\frac{2\mu}{\gamma}x\right)\cos\left(\frac{2\mu}{\gamma}x\right)\right],
$$
where si(.) et ci(.) are the sine integral and cosine integral functions \cite{Gradshteyn1965}. Expressed as a series around $x=0$ one obtains:
 \begin{eqnarray}
   v_{,x}(x,0)&=&-\frac{\alpha}{3\pi}\left(\frac{2\mu}{\gamma}x\right)^3\ln\left|\frac{2\mu}{\gamma}x\right|+\frac{2\alpha }{\pi}\left(\frac{2\mu}{\gamma}x\right)\ln\left|\frac{2\mu}{\gamma}x\right|+\alpha \mbox{sgn}\left(\frac{2\mu}{\gamma}x\right)+\frac{2\alpha}{\pi}(\Gamma-1)\left(\frac{2\mu}{\gamma}x\right) \nonumber\\
&&-\frac{\alpha\mbox{sgn}\left(\frac{2\mu}{\gamma}x\right)}{2}\left(\frac{2\mu}{\gamma}x\right)^2+\frac{\alpha}{3\pi}\left(\frac{11}{6}-\Gamma\right)\left(\frac{2\mu}{\gamma}x\right)^3+\cdots \nonumber
\end{eqnarray} 
where  $\Gamma$ is the Euler-Mascheroni constant and $\mbox{sgn}(x)=1$ for $x>0$, $\mbox{sgn}(x)=-1$ for $x<0$. The integration with respect to $x$ yields to (Eq. \ref{eqn : v}):
\begin{eqnarray}
v(x,0)-v(0,0)&=&\frac{\gamma \alpha }{2\mu}\left\{\left|\frac{2\mu}{\gamma}x\right|+\frac{1}{\pi}\left(\frac{2\mu}{\gamma}x\right)^2\ln\left|\frac{2\mu}{\gamma}x\right|+\frac{1}{\pi}\left(\Gamma-\frac{1}{2}\right)\left(\frac{2\mu}{\gamma}x\right)^2-\frac{1}{6}\left|\frac{2\mu}{\gamma}x\right|^3 \right. \nonumber\\
&&-\left. \frac{1}{12\pi}\left(\frac{2\mu}{\gamma}x\right)^4\ln\left|\frac{2\mu}{\gamma}x\right|+\frac{1}{12\pi}\left(\frac{19}{12}-\Gamma\right)\left(\frac{2\mu}{\gamma}x\right)^4+\cdots \right\}. \nonumber
\end{eqnarray}
Since $Y(x, 0)=-\alpha|x|+v(x,0)$ one obtains the asymptotic development of the shape of the surface for $x \ll \frac{\gamma}{\mu}$ (see Fig. \ref{fig : surface deformee}):

\begin{figure}[!h]
\includegraphics[width=0.5\textwidth]{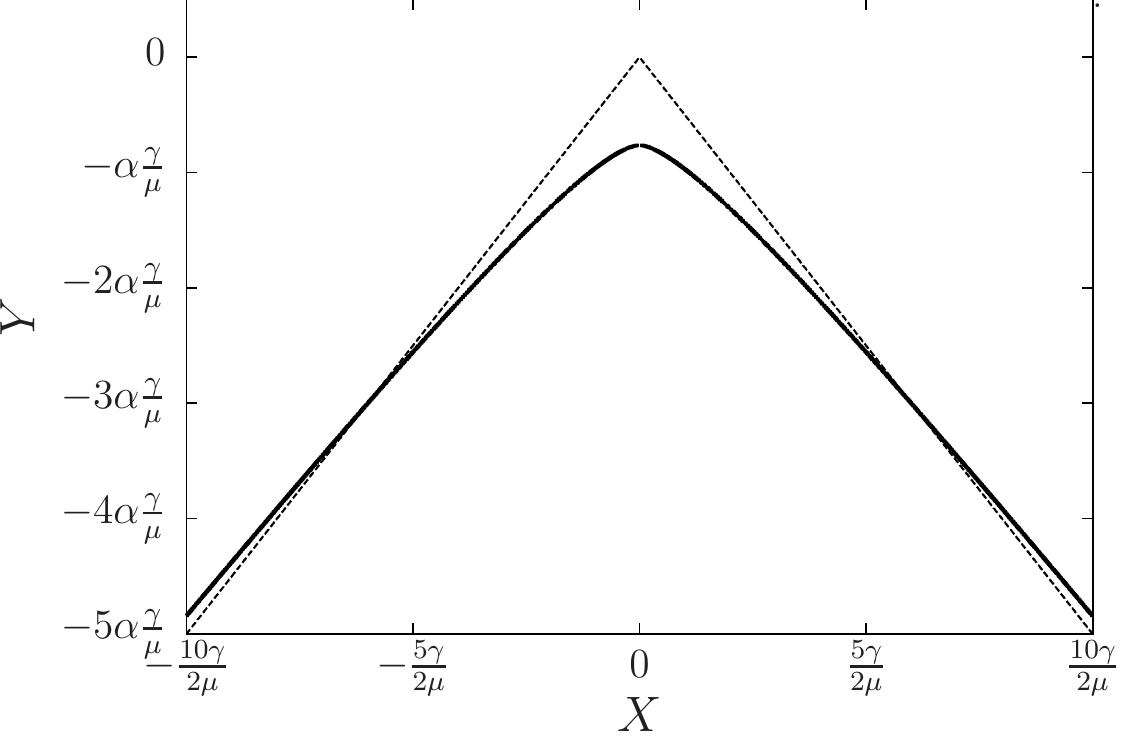}
\caption{Dashed line: wedge boundary without surface tension. Bold solid line: deformed boundary, numerically computed from Eqs. \ref{eqn : vx} and \ref{eqn : v}, with $d=\frac{10\gamma}{\mu}$.} \label{fig : surface deformee}
\end{figure}

\begin{equation}
Y(x)=\frac{\gamma}{2\mu}\alpha\left\{-\frac{2}{\pi}\ln\left(\frac{2\mu d}{\gamma} \right)+\frac{1}{\pi}\left(\frac{2\mu}{\gamma}x\right)^2\ln\left|\frac{2\mu}{\gamma}x\right|+\frac{1}{\pi}\left(\Gamma-\frac{1}{2}\right)\left(\frac{2\mu}{\gamma}x\right)^2-\frac{1}{6}\left|\frac{2\mu}{\gamma}x\right|^3+\cdots\right\},
\end{equation}
where $x$ can be replaced by $X$ at the linear order in $\alpha$ (from Eq. \ref{eq:uinteg}).
The curvature of the surface for $X \ll \frac{\gamma}{\mu}$ is
\begin{equation}
Y_{,XX}=\frac{4\alpha \mu}{\pi \gamma}\left\{\ln\left|\frac{2\mu}{\gamma}x\right|+\Gamma-2+\cdots\right\}.
\end{equation} 
It diverges logarithmically near $x =0$.  

\subsection{Smoothening of a sharp angle by surface tension}
\label{sec:sharp}
This is the  opposite limit, namely the one of an initially sharp angle smoothened by surface tension. This situation is analyzed in the lubrication limit, namely by assuming that all quantities change much more rapidly as a function of the variable perpendicular to the axis of symmetry than with respect to the position along this axis (longitudinal variable). Therefore, at leading order, one neglects the dependence with respect to the longitudinal variable and solves the problem of straining of a long rectangular piece of elastic material under the effect of surface tension along its long side. Afterwards this "base" solution is used to solve the wedge problem by assuming slow changes of this solution with respect to the longitudinal position. 

\subsubsection{Strain induced by capillary action on a long elastic rectangle} \label{sec : contraction}

We consider a long flat parallelepipedic ribbon of dimension 
$D\times L \times l$ with$D\gg L \gg l$ (Fig. \ref{fig : ruban1}). 
It is made of an incompressible neo-Hookean material of shear modulus $\mu$. It is put in a medium such that its surface tension is $\gamma$, which will strain it. 
One assumes an homogeneous strain, equivalent to neglect the rounding of the sides, something of second order with respect to the small parameters $L/D$ and $l/L$.  
Under the stress brought by surface tension, the length of the sides of the ribbon change to 
$\lambda_3 D \times \lambda_2 L \times \lambda_1 l$, where $\lambda_{1,2,3}$ are numbers to be found. Volume conservation imposes $\lambda_1 \lambda_2  \lambda_3=1$. The values of the $lambda$'s are found by minimizing the total energy of the strained system arises from the strain in volume (${\cal E}_{bulk}$)  and from  the change of area (${\cal E}_{surf}$):
\begin{figure}[!h]
\includegraphics[width=0.5\textwidth]{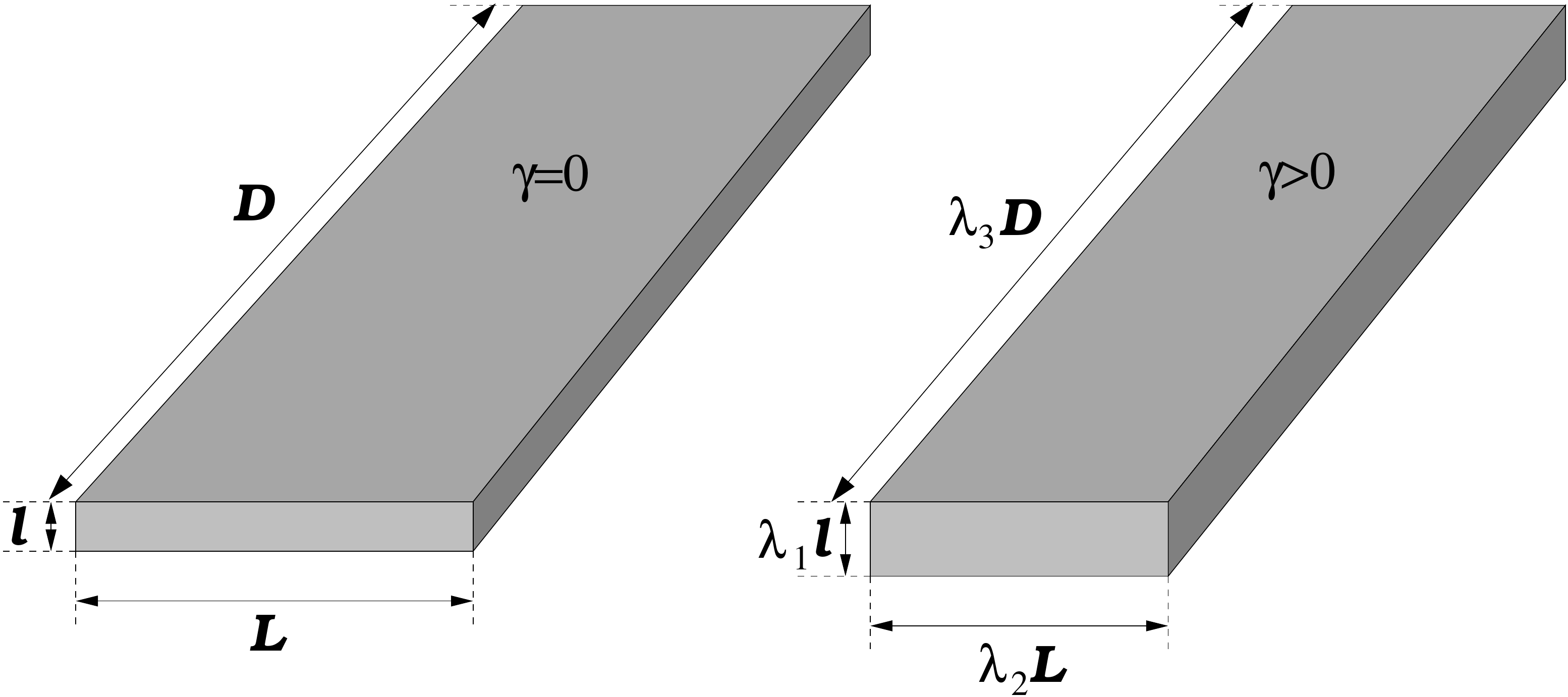}
\caption{Long and flat parallelepipedic elastic ribbon deformed by its surface tension.} \label{fig : ruban1}
\end{figure}

$${\cal E}_{bulk}=\frac{\mu}{2}\left(\lambda_1^2+\lambda_2^2 + \lambda_3^2\right)LlD,$$
and
$${\cal E}_{surf}=2\gamma (\lambda_1\lambda_2Ll+ \gamma \lambda_2\lambda_3 DL+\lambda_1\lambda_3 Dl)\simeq 2\gamma \lambda_2\lambda_3 DL.$$
The approximation for the capillary energy is valid in the limit $D\gg L \gg l$. 
Taking into account the conservation of volume $\lambda_3=1/(\lambda_1\lambda_2)$ the energy of the deformed state becomes:
$$
{\cal E}_{bulk}+{\cal E}_{surf}=\frac{\mu}{2}\left(\lambda_1^2+\lambda_2^2 + \frac{1}{\lambda^2_1\lambda^2_2}\right)LlD+\frac{2\gamma}{\lambda_1}.
$$

The values of $(\lambda_1, \lambda_2)$ at equilibrium are found by minimizing the total energy, leading to  
$\lambda_1=\left(1+\frac{2\gamma}{\mu l} \right)^{1/3}$ and 
$\lambda_2=\lambda_3=\left(1+\frac{2\gamma}{\mu l} \right)^{-1/6}$.

\subsubsection{Capillary roughening of a wedge in the small angle limit}  
\begin{figure}[!h]
\includegraphics[width=0.3\textwidth]{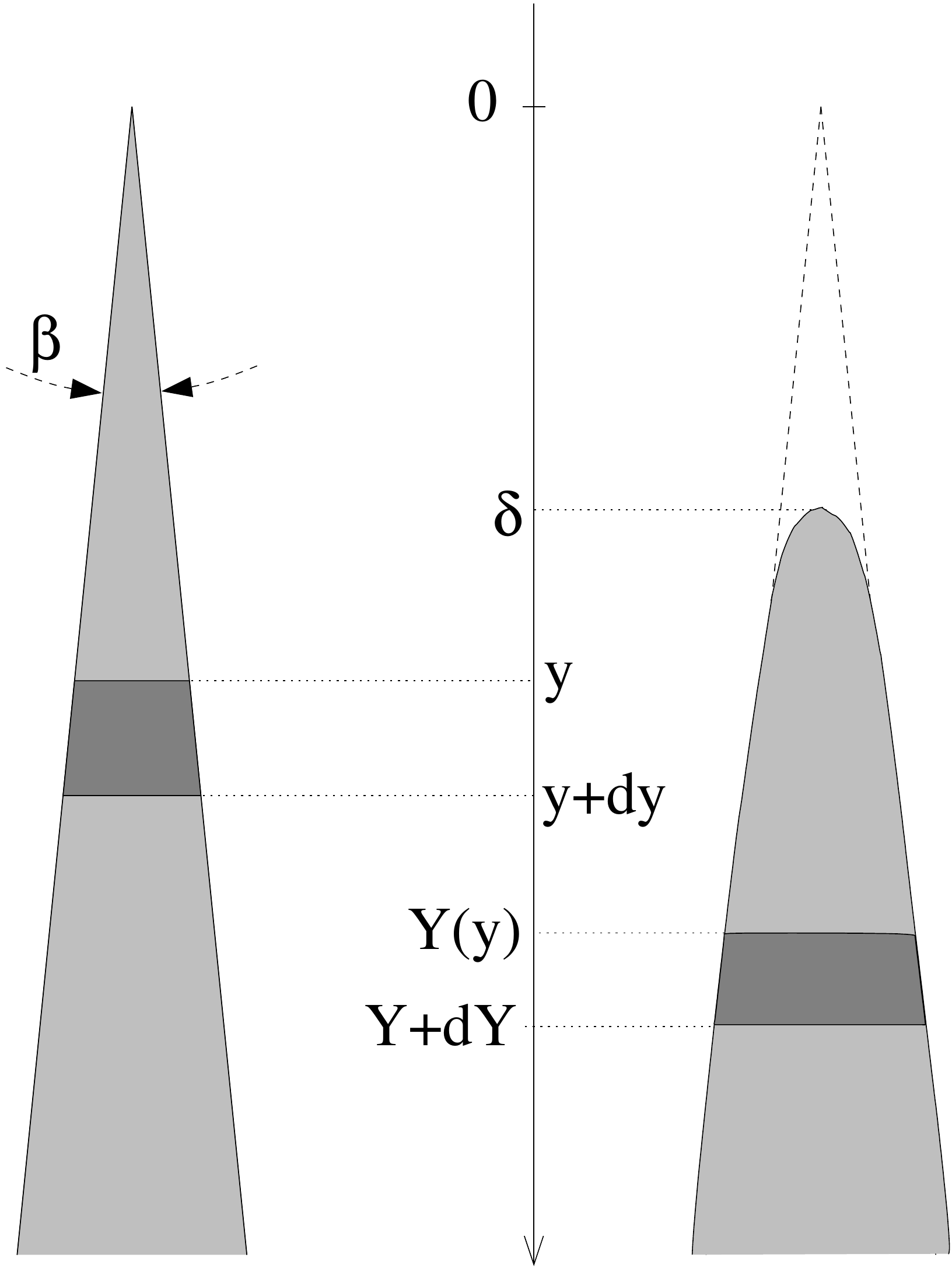}
\caption{Capillary driven deformation of a two dimensional wedge of angle $\beta \ll 1$. For the calculation, it is divided into slices of height d$y$.} \label{fig : wedge ruban}
\end{figure}

We consider below the change of shape of a very sharp wedge of angle 
$\beta\ll 1$, assumed to be very long in the axial direction. This wedge will be also assumed to be made of incompressible neo-Hookean material of shear modulus $\mu$. We assume that this wedge is prepared in a stress-free state without surface tension, and that it is put into a liquid which changes the surface tension from zero to $\gamma$. The slice of material located initially between  $y$ and $y+\mbox{d}y$ is located, in the deformed state between $Y$ and $Y+\mbox{d}Y$ with $\mbox{d}Y=\lambda(y) \mbox{d}y$ where $\lambda(y)=\left(1+\frac{2\gamma}{\mu \beta y} \right)^{-1/6}$(see Fig. \ref{fig : wedge ruban}). The local contraction factor is just the one computed before for a long and thin parallelepiped (Sec. \ref{sec : contraction}). It is a function of the ratio of the elasto-capillary length to the width of the wedge at $y$, namely $\beta y$. The limit $\beta\ll 1$ ensures that the elastic stress at the contact surface between two adjacent slices is negligible compared to the capillary stress acting on the free surface. 

Let $d >0$ be the length scale such that, for $y=d$, the $y$-component of the displacement is zero, it is such that $Y(d)=d$. One can think that the wedge is initially put on a fixed plate of abscissa $y=d$, and that it can slide freely on this plate. Thereafter it will be possible to take the limit $d$ infinite. 

From $\mbox{d}Y=\lambda(y)\mbox{d}y$ one derives by integration:  
$$
Y(y)-Y(d)=\int_d^y\lambda(y')\mbox{d}y'=\int_d^y\lambda(y)\mbox{d}y=\frac{2\gamma}{\mu\beta}\int_{\frac{\mu \beta d}{2\gamma}}^{\frac{\mu \beta y}{2\gamma}}\left(1+\frac{1}{t}\right)^{1/6}\mbox{d}t=\frac{2\gamma}{\mu \beta}\left[f\left(\frac{\mu \beta y}{2\gamma}\right)-f\left(\frac{\mu \beta d}{2\gamma}\right)\right],
$$
where the function $f(t)$ is defined as
$$ f(t) = \int_0^t \left(1 + \frac{1}{t'}\right)^{1/6} {\mbox{d}} t' \mathrm{.}$$

From the definition, $f(0)=0$ and $f(t)\sim t$ for $t$ large.

Therefore:
\begin{equation}
Y(y)=\frac{2\gamma}{\mu \beta}f\left(\frac{\mu \beta y}{2\gamma}\right)+d\left[1-\left(\frac{2\gamma}{\mu \beta d}\right)f\left(\frac{\mu \beta d}{2\gamma}\right)\right].
\label{eqn : wedge analytic total}
\end{equation}

Since $f(t) \simeq t-\frac{1}{6}\ln t-0.126+\cdots$ for $t\gg 1$, one can simplify Eq. \ref{eqn : wedge analytic total} for $d\gg \frac{\gamma}{\beta \mu}$

\begin{equation}
Y(y)=\frac{2\gamma}{\mu \beta}f\left(\frac{\mu\beta y}{2\gamma}\right)+\frac{\gamma}{3\mu\beta}\ln\left(\frac{\mu \beta d}{2\gamma} \right).
\end{equation}

\begin{figure}[!h]
\includegraphics[width=0.5\textwidth]{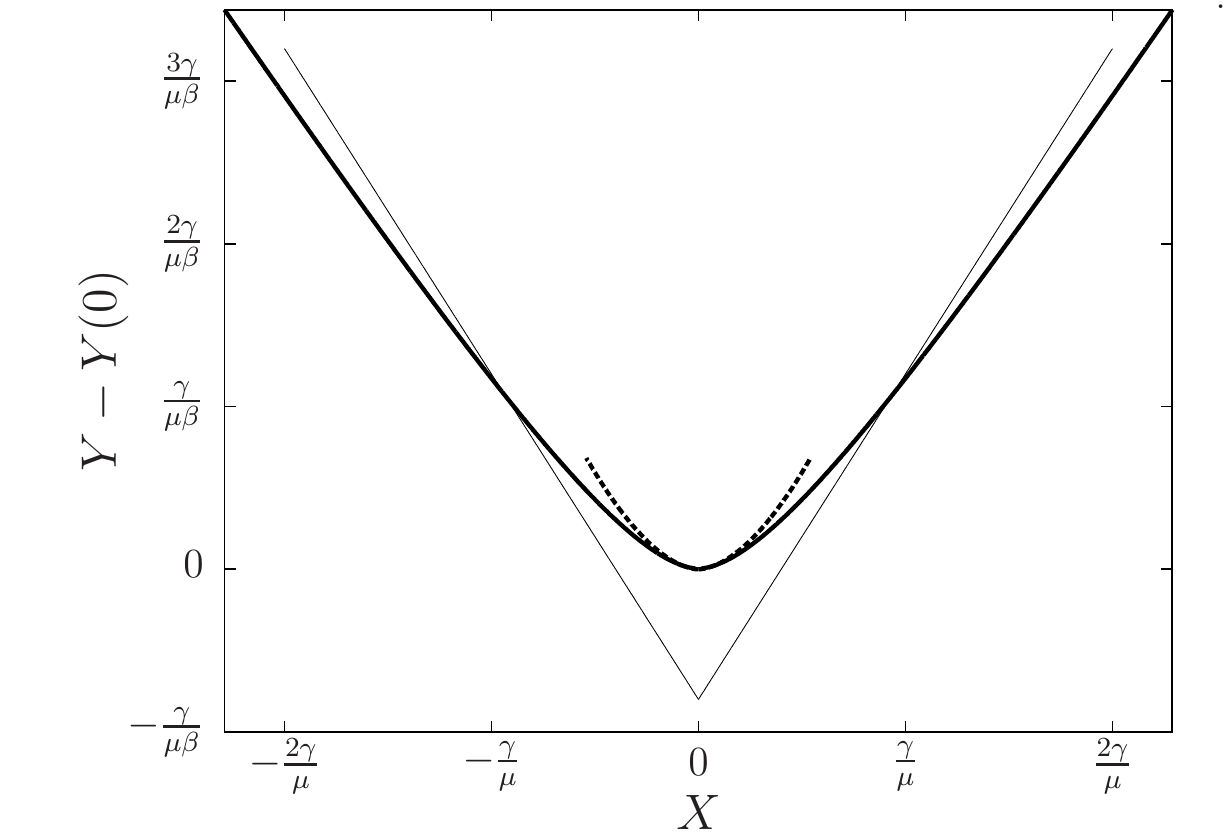}
\caption{Thin solid line: wedge boundary without surface tension. Bold solid line: deformed boundary, computed from Eq. \ref{eqn : parametrique wedge}. Dashed line: asymptotic expression Eq. \ref{eqn : wedge premier ordre} for  $\mu \beta y \ll \gamma$.} \label{fig : wedge}
\end{figure}

Moreover the width of the slice located in between $y$ and $y+\mbox{d}y$ is $2x=\beta y$. Upon deformation this width becomes $2X=\beta y/\lambda^2(y)$, namely:
$$
2 X=\beta y\left(1+\frac{2\gamma}{\mu \beta y} \right)^{1/3}=\left(\frac{2\gamma}{\mu}\right)\left(1+\frac{2\gamma}{\mu \beta y}\right)^{1/3}\left(\frac{\mu \beta y}{2\gamma}\right) \mathrm{.}
$$
The shape of the surface resulting from the deformation of the wedge by surface tension is given by the following set of two parametric equations (with  $t=\frac{\mu \beta y}{2\gamma}\in [0,\infty]$):
\begin{equation}
\left\{ \begin{array}{l}
X(t)=\frac{\gamma}{\mu}\times t\times \left(1+\frac{1}{t}\right)^{1/3},\\
Y(t)=\frac{\gamma}{\beta \mu} \times 2 \times f(t)+\frac{\gamma}{3\mu \beta}\ln\left(\frac{\mu \beta d}{2\gamma} \right).
\end{array} \right.
\label{eqn : parametrique wedge}
\end{equation}
The curve so defined is plotted on Fig. \ref{fig : wedge}. Since $Y\propto 1/\beta$, the strain increases as $\beta$ decreases. This is because the ratio of the free surface to the volume is larger as $\beta$ is smaller.

The displacement $\delta$ of the tip is along the axis of symmetry of the wedge: $\delta=Y(0)$. From Eq. \ref{eqn : parametrique wedge} one finds: 
$$
\delta=\frac{\gamma}{3\mu \beta}\ln\left(\frac{\mu \beta d}{2\gamma} \right).
$$
Because it diverges at large distances from the tip, if one assumes a wedge extending to infinity ($d\ll \gamma/\mu$), $\delta$ depends logarithmically of the full size of the object.\\

Assuming $\frac{\mu \beta y}{2\gamma} \ll 1$, which defines the neighborhood of the tip, one finds for the local behavior near the deformed tip, in the parametric representation:  
$$
Y=\delta+\frac{12\gamma}{7\mu \beta}\left(\frac{\mu \beta y}{2\gamma}\right)^{7/6}+ \cdots
$$
and 
$$
X=\frac{\gamma}{\mu}\left(\frac{\mu \beta y}{2\gamma} \right)^{2/3}.
$$
Therefore:
\begin{equation}
Y=\delta +\frac{12\gamma}{7\mu\beta}\left(\frac{ \mu X}{\gamma} \right)^{7/4}+\cdots
\label{eqn : wedge premier ordre}
\end{equation}
This shows, as expected, a non-smooth behavior near the deformed tip, but with a singularity far weaker than the one of the undeformed tip. This one has a singular slope, although after deformation, the curvature is singular, but the slope is not. 

\section{Conclusion}
\label{sec:conclusion}
We have shown the smoothening effect of surface tension in soft solids obeying the equations of neo-Hookean elasticity. Specifically we focused on the opposite limits of shallow wedges and of sharp edges, where analytical results can be obtained. In both cases, the displacement of the tip due to surface tension depends on a large scale cut-off via a logarithm, likely not very easy to put in evidence experimentally. Perhaps it would be easier to put in evidence the singularity of the curvature at the tip, when surface tension is turned on. Notice that, although in both cases (sharp and shallow) one finds a singularity of the curvature, the exponents are different (the logarithm found for the shallow wedge can be taken as a zero exponent). An obvious conjecture, perhaps answerable by numerical studies, is that the exponents of this singularity of curvature depend on the angle of the tip in the rest state of the edge. We notice also that this roughening of the edges by surface tension has been observed \cite{Mora2013} recently in experiments on soft hydrogels of finite lateral size, with a good agreement between the experiments and the results of computer studies based on the neo-Hookean model applied to the case of a wedge of finite angle $\alpha=90^o$.  


\end{document}